\documentclass[final,twocolumn]{elsarticle}

\usepackage[T1]{fontenc}
\usepackage[utf8]{inputenc}
\usepackage{amsmath}
\usepackage{amssymb}
\usepackage{graphicx}
\usepackage{booktabs}
\usepackage{hyperref}
\usepackage{url}

\journal{Journal of Network and Computer Applications}

\begin{document}

\begin{frontmatter}

\title{ZK-Rollup for Hyperledger Fabric: Architecture and Performance Evaluation}

\author[inst1]{Sania Siddiqui}
\author[inst1]{Neha\corref{cor1}}
\cortext[cor1]{Corresponding author}
\ead{p20240030@pilani.bits-pilani.ac.in}
\author[inst1]{Hari Babu K}

\address[inst1]{Birla Institute of Technology and Science, Department of Computer Science, Pilani, India}

\begin{abstract}
A big challenge posed in blockchain-centric platforms is achieving scalability while also preserving user privacy. This report details the design, implementation and evaluation of a Layer-2 scaling solution for Hyperledger Fabric using Zero-Knowledge Rollups (ZK-Rollups). The proposed architecture introduces an off-chain sequencer that accepts transactions immediately and sends them for batching into a Merkle tree based rollup, using ZK proofs to attest to the correctness and verifiability of the entire batch. 

The design aims to decouple transaction ingestion from actual on-chain settlements to address Fabric’s scalability limitations and increase throughput under high load conditions. The baseline architecture in Hyperledger Fabric constrains the transaction requests due to the process of endorsement, ordering and validation phases, leading to a throughput of 5-7 TPS with an average latency of 4 s. Our Layer-2 solution achieves an ingestion throughput of 70-100 TPS, leading to an increase of nearly 10x due to the sequencer’s immediate acceptance of each transaction and reducing the client-side perceived latency by nearly 80\% to 700-1000 ms. This work demonstrates that integrating ZK-Rollups solution in Hyperledger Fabric enhances scalability while not compromising on the guarantee of security that a permissioned blockchain network inherently has.
\end{abstract}

\end {frontmatter}

\section{Introduction}

Blockchain systems have gained widespread adoption across various enterprises due to their immutability, auditability, and decentralization. The issue with public blockchain implementations is the lack of sufficient privacy controls for sensitive case data. Public blockchains have more immutable transactions and are secure due to the use of high consensus algorithms~\cite{dai2019blockchain}, but the issue of scalability and lower throughput arises, and thus leads to high latency for recording transactions in a distributed network~\cite{coyne2017blockchains,dai2017toward}.

We see that there exists a need for a framework that allows the building of scalable and privacy-preserving blockchain platforms. We face two big challenges. Firstly, the assets that are going to be stored might be large and complex. Trying to store all of them directly on-chain to a blockchain ledger is computationally inefficient. Secondly, users must be able to prove attributes about themselves (e.g., age verification, ownership of an asset) without revealing sensitive underlying data ~\cite{raipurkar2023digital}.

In this paper, we attempt to resolve some of those challenges. Hyperledger Fabric, being one of the most widespread permissioned blockchain networks, faces well-documented scalability challenges under high transaction loads due to its endorsement, ordering and validation of every transaction which results in throughput bottlenecks and increased latency as network demand increases~\cite{androulaki2018fabric,vukolic2015quest}.

Although ZK-Rollups have been widely studied in public account based blockchains such as Ethereum ~\cite{braun2018zkrollups}, their application in permissioned, enterprise-oriented blockchains has been underexplored. Hyperledger Fabric lacks the support for batching or verification based on Merkle-root rollups, which leaves open an opportunity to enhance the workflow of Hyperledger Fabric by integrating a Layer-2 rollup model that is tailored to its modular transaction flow and endorsement model. 

In this work, we implement a Layer-2 rollup model for Hyperledger Fabric which uses a Poseidon-based 5-level Merkle tree and a Permutations over Lagrange-Bases for Oecumenical Non-interactive arguments of Knowledge (PLONK) based zero-knowledge system to increase perceived throughput and latency. The sequencer batches each transaction into groups of 32 and generates a proof of correctness, also making the Fabric system more scalable by only storing the batch metadata directly on-chain on Fabric via a custom chaincode function. Section II describes the architecture of the network and methodology used to construct the ZK-Rollup. Section III discusses the experimental setup used to compare the direct on-chain transactions with the off-chain ingestion pipeline. We discuss our results in Section IV and talk about related work that has proposed the use of ZKPs on Hyperledger Fabric network and their findings in Section V. Section VI presents the conclusion and discussion from our findings along with possible future work.

The system consists of three primary layers:
\begin{enumerate}
    \item \textbf{Client Layer:} Users interact with assets (create, transfer, and view) through a REST API exposed by the sequencer.
    \item \textbf{Layer-2 (Off-chain):} A sequencer service that queues incoming transactions, batches them into groups of 32, and generates zero-knowledge proofs.
    \item \textbf{Layer-1 (On-chain):} The Hyperledger Fabric network verifies the ZK-proof and stores the batch commitment in the form of the Merkle root.
\end{enumerate}

\section{System Architecture and Methodology}

This section describes the framework of the system architecture used and the methodology of the ZK-Rollup architecture along with algorithms used to implement the sequencer and create the zero knowledge proofs. 

To improve upon the bottleneck of a direct on-chain proof verification process, we introduce a computation layer that makes sure that the transactions are ingested faster. The sequencer acts as a Layer-2 off-chain computation engine. It receives asset creation requests from client applications, stores them temporarily in Redis and periodically batches them into groups of 32 transactions. A background loop then converts them into Merkle tree and generates a ZK-snark proof to attest to the correctness of the Merkle root. 

\begin{figure*}[htbp]
    \centering
    \includegraphics[width=\textwidth]{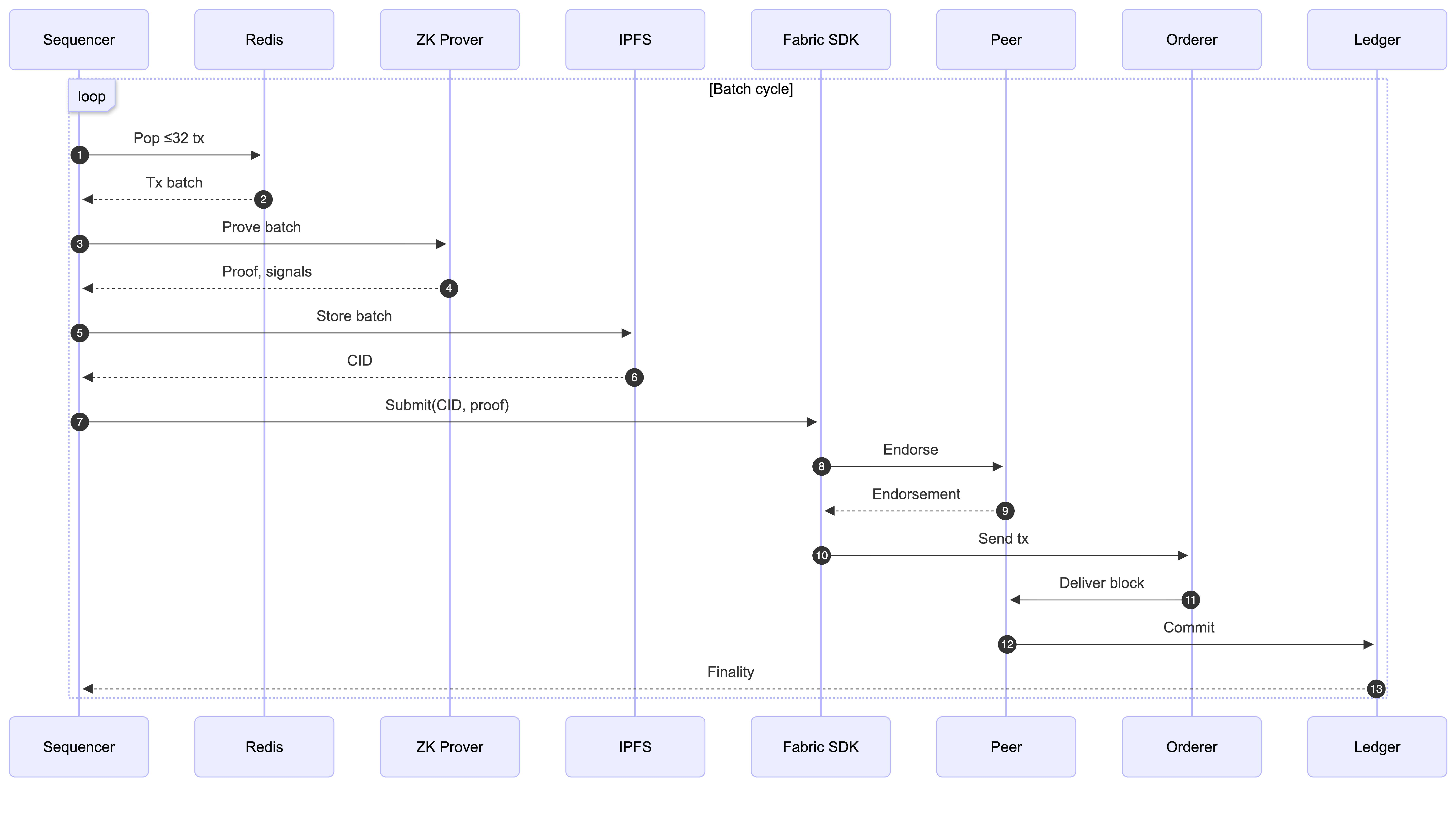}
    \caption{Execution flow of the system architecture}
    \label{fig:architecture}
\end{figure*}

Hyperledger Fabric finally only receives the proof, Merkle root, InterPlanetary File System (IPFS) Content Identifier (CID) and metadata, thus reducing the blockchain’s computational overhead. Figure~\ref{fig:architecture} presents a high-level sequence diagram illustrating the overall execution flow of the proposed system architecture.

\subsection{Blockchain Network}

\begin{figure*}[htbp]
    \centering
    \includegraphics[width=\textwidth]{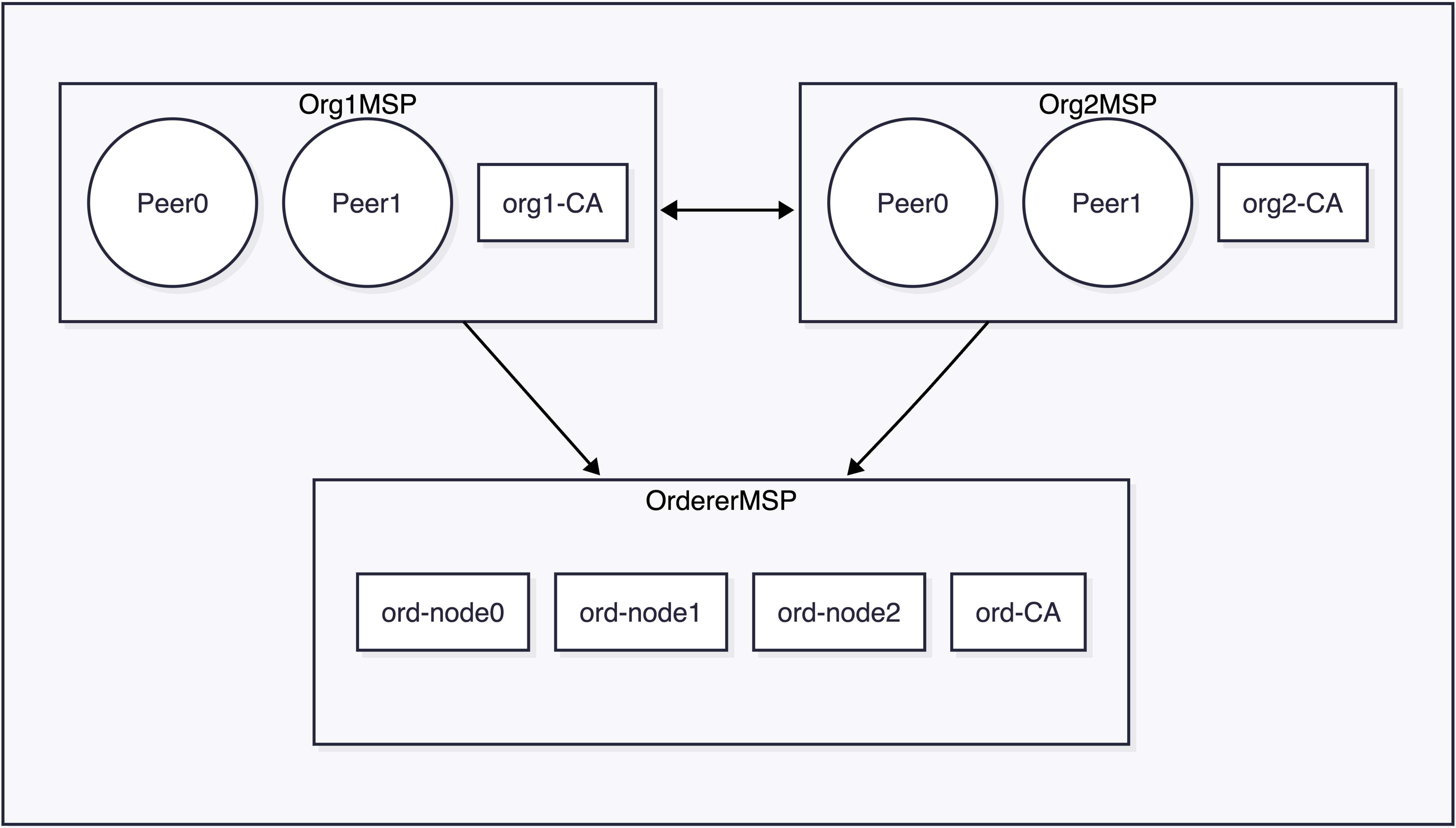}
    \caption{The implemented two-organization Hyperledger Fabric network}
    \label{fig:fabric_network}
\end{figure*}

We built a Hyperledger Fabric Network deployed in a Kubernetes cluster, which consists of two separate organizations with two peers each to execute smart contracts and verify chaincode transactions, as illustrated in Figure~\ref{fig:fabric_network}. Each peer nodes maintain a local copy of the distributed ledger. The network also includes a three-node ordering service configured with the RAFT consensus protocol to ensure fault-tolerance.

A demo channel connects all peers from both the organisations, creating a private subnet to enable secure communication and private blockchain transactions. A Certification Authority also exists in each organisation to manage and issue two kinds of digital certificates in the Hyperledger Fabric Network: Signing certificates (identity certificates) and Transport Layer Security (TLS) certificates (encrypting all communication between the nodes such as P2P, P2O, etc).

\subsection{Zero-Knowledge Rollup}

Clients interact with the sequencer through a REST API and send transaction requests. Each transaction consists of the \texttt{CreateAsset} chaincode function which consists of the AssetID, Participant Identifier and an IPFS CID referencing the original asset. The sequencer validates each transaction and sends it to a Redis-backed transaction pool. 

The sequencer periodically checks the transaction pool and extracts up to 32 transactions. In the case of fewer available transactions, it pads the batch with dummy transactions so as to keep a consistent circuit size, maintaining efficient zero-knowledge proof generation. 

The Zero-Knowledge Rollup here works by separating the transaction workflow into two parts: \emph{ingestion} and \emph{settlement}. The process of Ingestion is fast as the high-speed sequencer immediately accepts the transactions from the user as they come in, putting them in a Redis queue - thus making for an instant response and thus low latency. In the background, the process of settlement runs paralelly as the sequencer aggregates 32 transactions into one singular zero-knowledge proof, using a Merkle tree of depth five, as shown in Algorithm~\hyperlink{alg:merkle}{1}. 

This algorithm produces a Merkle root that serves as a cryptographic commitment to the entire batch. It is deterministic and ensures that the same ordered set of transactions will always produce the same Merkle root. The number of hash computations are constant, thus ensuring that the proving costs are always bounded. A PLONK based zero-knowledge Succinct Non-Interactive Argument of Knowledge (ZK-SNARK) circuit is used to prove the correctness of the Merkle root computed from the hashes. The blockchain can then verify the correctness of an entire batch of 32 transactions while only using the single succinct verification of the Merkle root, significantly reducing on-chain computational overhead.

\bigskip
\hypertarget{alg:merkle}{}

\begin{center}
\small
\textbf{Algorithm 1. Poseidon-based Merkle Tree Construction for 32 Leaves}

\vspace{0.5em}

\begin{verbatim}
Input: leaves[0..31]
Output: root

Define HashLeftRight(x, y) = Poseidon(x, y)

for i = 0 to 15:
    level1[i] = HashLeftRight(leaves[2i], leaves[2i+1])

for i = 0 to 7:
    level2[i] = HashLeftRight(level1[2i], level1[2i+1])

for i = 0 to 3:
    level3[i] = HashLeftRight(level2[2i], level2[2i+1])

for i = 0 to 1:
    level4[i] = HashLeftRight(level3[2i], level3[2i+1])

root = HashLeftRight(level4[0], level4[1])
return root
\end{verbatim}
\end{center}

\bigskip

\subsection{IPFS and On-Chain Submission to Hyperledger Fabric}

All non-dummy transactions within each batch are serialized into JSON format and uploaded to the InterPlanetary File System (IPFS). The resulting IPFS Content Identifier (CID) uniquely references the batch data, enabling efficient off-chain storage through immutable content addressing. This design minimizes on-chain state growth while ensuring data integrity and availability.

\begin{figure*}[htbp]
    \centering
    \includegraphics[width=\textwidth]{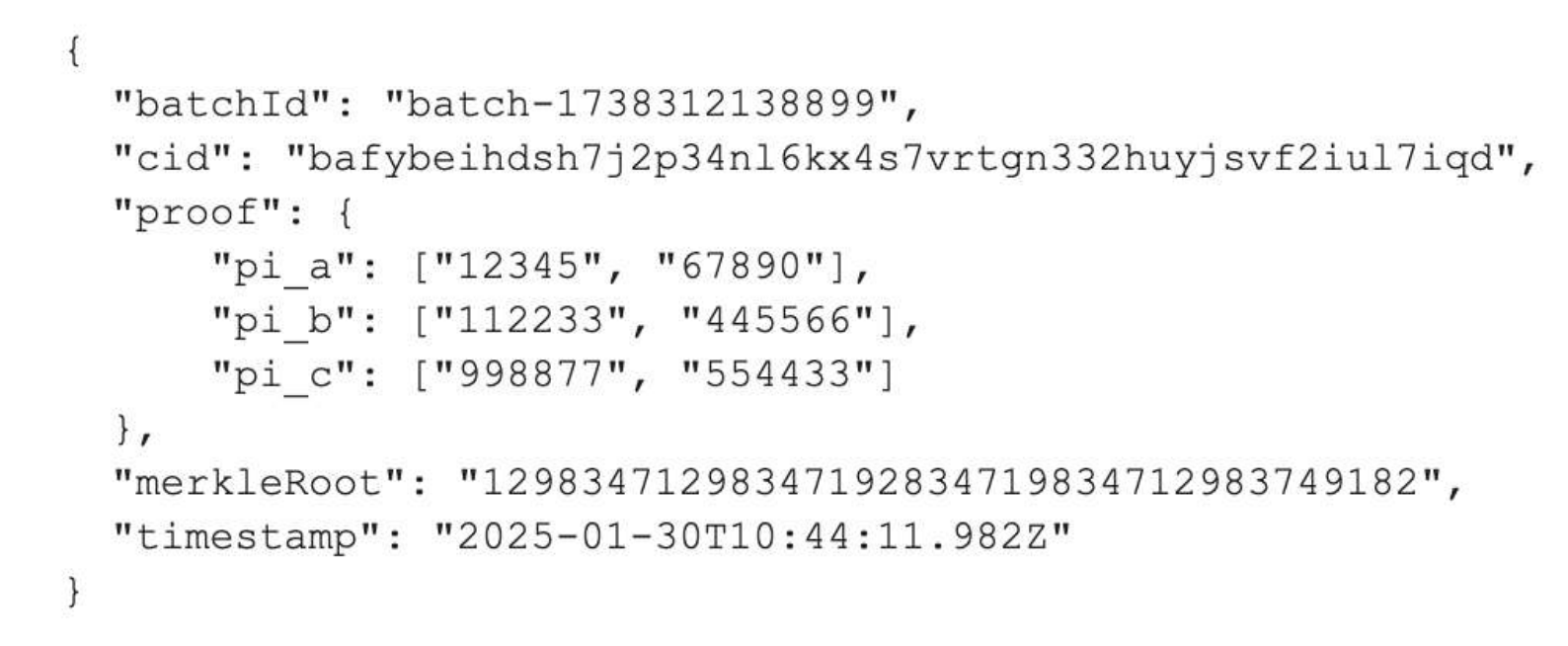}
    \caption{On-chain metadata stored in Hyperledger Fabric}
    \label{fig:onchain_metadata}
\end{figure*}

Once the zero-knowledge proof is generated, the sequencer submits the proof, the corresponding Merkle root, the IPFS CID, and associated metadata to the Hyperledger Fabric network. The chaincode processes and records this metadata on-chain, thereby immutably persisting the batch commitment without storing individual transaction details. Figure~\ref{fig:onchain_metadata} illustrates the on-chain metadata structure.

\section{Experimental Setup}

This section details the hardware and software environment, experiment parameters and benchmarking tools and algorithms used to evaluate the proposed ZK-rollup architecture.

\subsection{Testbed Environment}

\begin{figure}[!t]
\centering
\includegraphics[width=\linewidth]{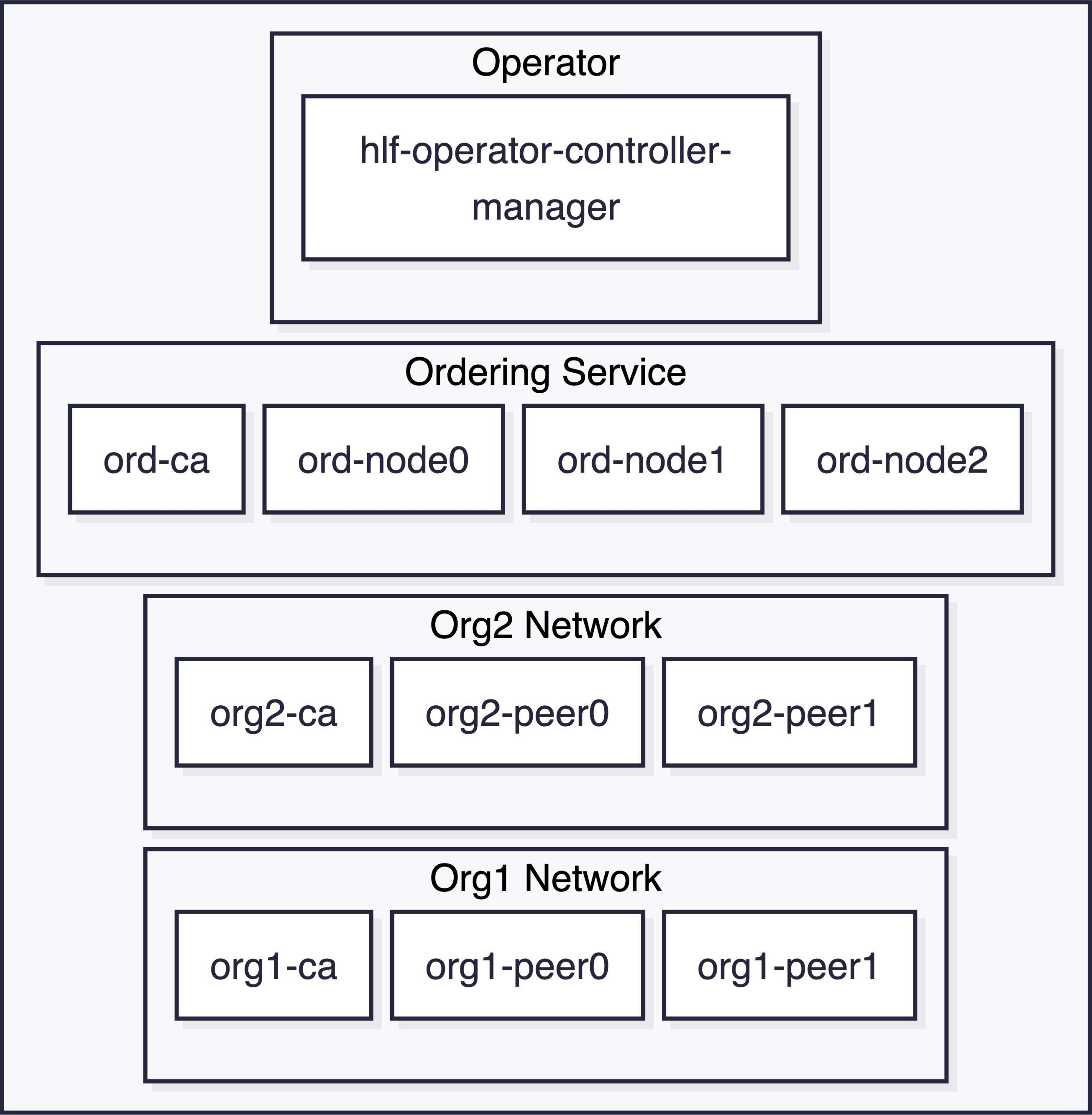}
\caption{Container overview of the Hyperledger Fabric network.}
\label{fig:fabric-core}
\end{figure}

All experiments were conducted on a local Kubernetes-in-Docker (KinD) cluster running on a local host machine with 8 CPU cores and 8 GB RAM. The KinD cluster deployed a 2 organization and 4 peers, Hyperledger Fabric architecture and is shown in Figure~\ref{fig:fabric-core} as a container.

\begin{figure}[!t]
\centering
\includegraphics[width=\linewidth]{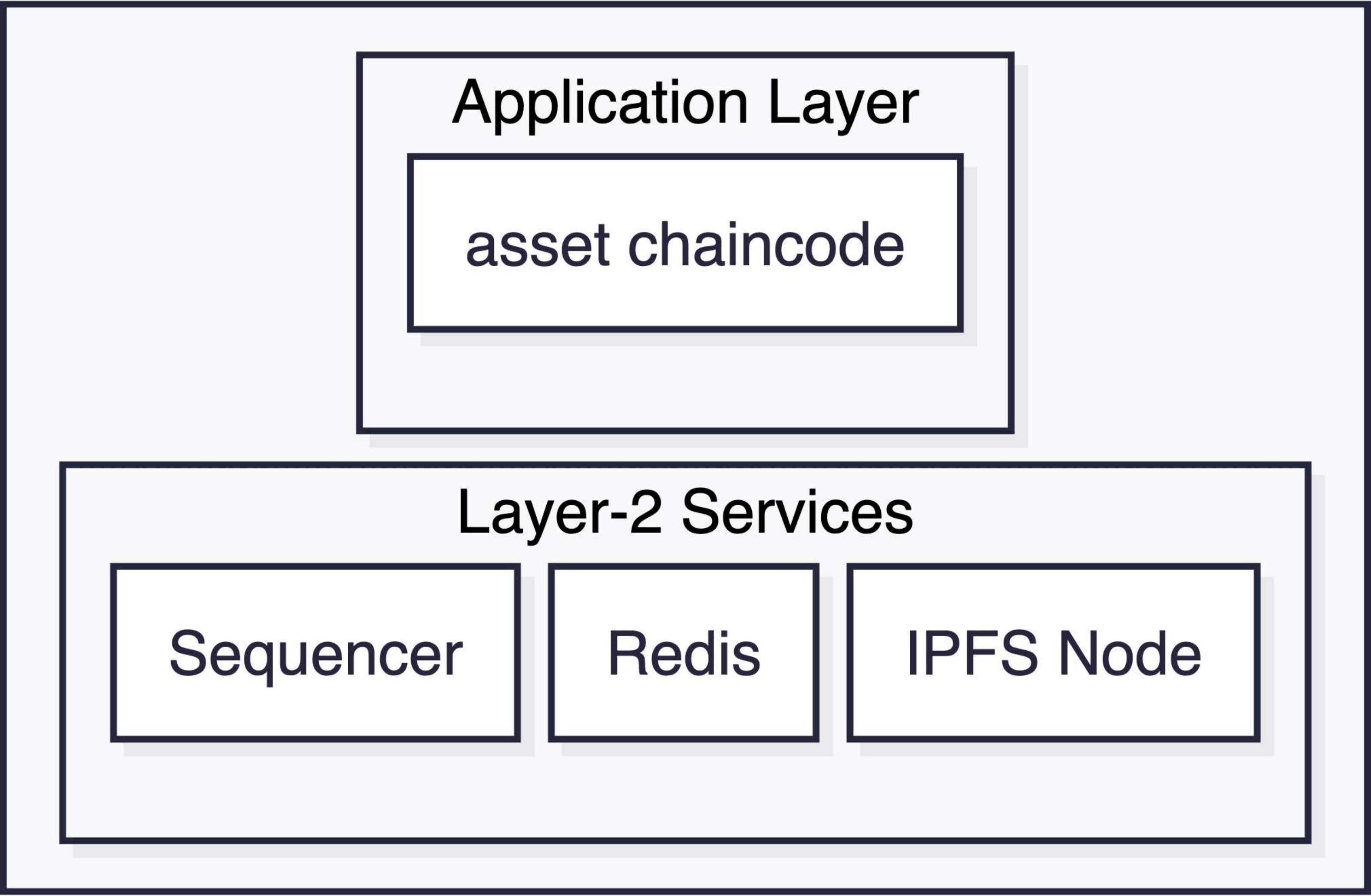}
\caption{Layer-2 services and application layer integrated with the Fabric network.}
\label{fig:layer2}
\end{figure}

Along with the components of the fabric network (orderers, peers and chaincode), the off-chain sequencer, redis transaction pool and IPFS node were also deployed separately in the same cluster and illustrated in Figure~\ref{fig:layer2}. 

The sequencer communicated with the Fabric network using the Fabric Gateway SDK and interfaced with IPFS through the \texttt{ipfs-http-client} library.

To assess the throughput and generate load, we used k6, executed from outside the cluster. All of the scripts were run for a fixed duration of 30 seconds under different virtual user configurations to stress both the baseline and the ZK-rollup architecture.

\subsection{Baseline Workload: Direct On-Chain transactions}

This test measures the performance of direct, synchronous on-chain transactions without any L-2 component. The baseline script uses 20 virtual users and sends direct HTTP POST requests to the client APIs of Org1 and Org2 for a total duration of 30 seconds, in turn invoking the \texttt{CreateAsset} chaincode function directly on Hyperledger Fabric. It follows the standard Hyperledger Fabric transaction flow, where every transaction request gets synchronously submitted to Fabric and must be endorsed by the relevant peers, then ordered by the ordering service and finally validated and committed to the ledger before the client receives a response. 

\begin{figure}[htbp]
    \centering
    \includegraphics[width=0.48\textwidth]{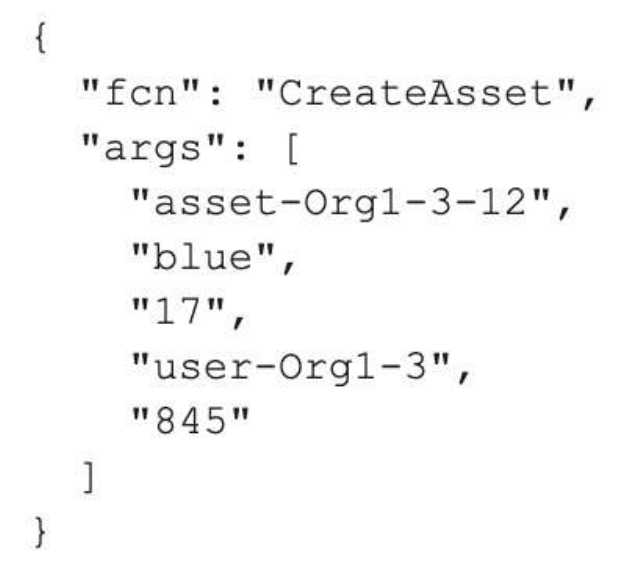}
    \caption{Randomised transaction value}
    \label{fig:random_tx}
\end{figure}

The script randomly selects either of the organization’s ports, and sends a request to the \texttt{/submit-direct} end point with a randomised asset (Figure~\ref{fig:random_tx}). Each request waits for an HTTP 200 OK to signify that the transaction has been fully endorsed, ordered and committed on-chain. k6 records the per-request latency and success rate. The overall throughput (request/s) is used as the baseline Fabric performance under the load. This design guarantees strong consistency and correctness, but limits throughput under high load, as clients must wait for the finalisation of the transaction before proceeding. 

\subsection{ZK-Rollup Workload: Off-chain sequencer ingestion throughput}

This script measures the ingestion capacity and also the client-perceived latency of the off-chain sequencer. It sends HTTP POST requests to the \texttt{/submit} endpoint exposed by client APIs of Org1 and Org2, which in turn forwards the requests to the sequencer. This configuration uses 50 virtual users (due to higher concurrency than the baseline model) for a total duration of 30 seconds. Each request waits for an HTTP 202 Accepted to indicate that the transaction has been accepted by the sequencer, not necessarily completing on-chain settlement. This captures the perceived ingestion throughput (request/s) and latency, decoupled from the slower Fabric consensus path.

\subsection{ZK Generation Time: True settlement throughput and generation of ZK-Proofs}

For the ZK-Rollup architecture, additional true settlement throughput and ZK-Proof generation time metrics are recorded using the sequencer logs. These record the time taken to generate a proof and commitment of the batch to Fabric. It allows for a distinction between ingestion capacity and on-chain settlement rate for the Layer-2 Architecture.

\section{Evaluation}

\begin{figure*}[htbp]
    \centering
    \includegraphics[width=\textwidth]{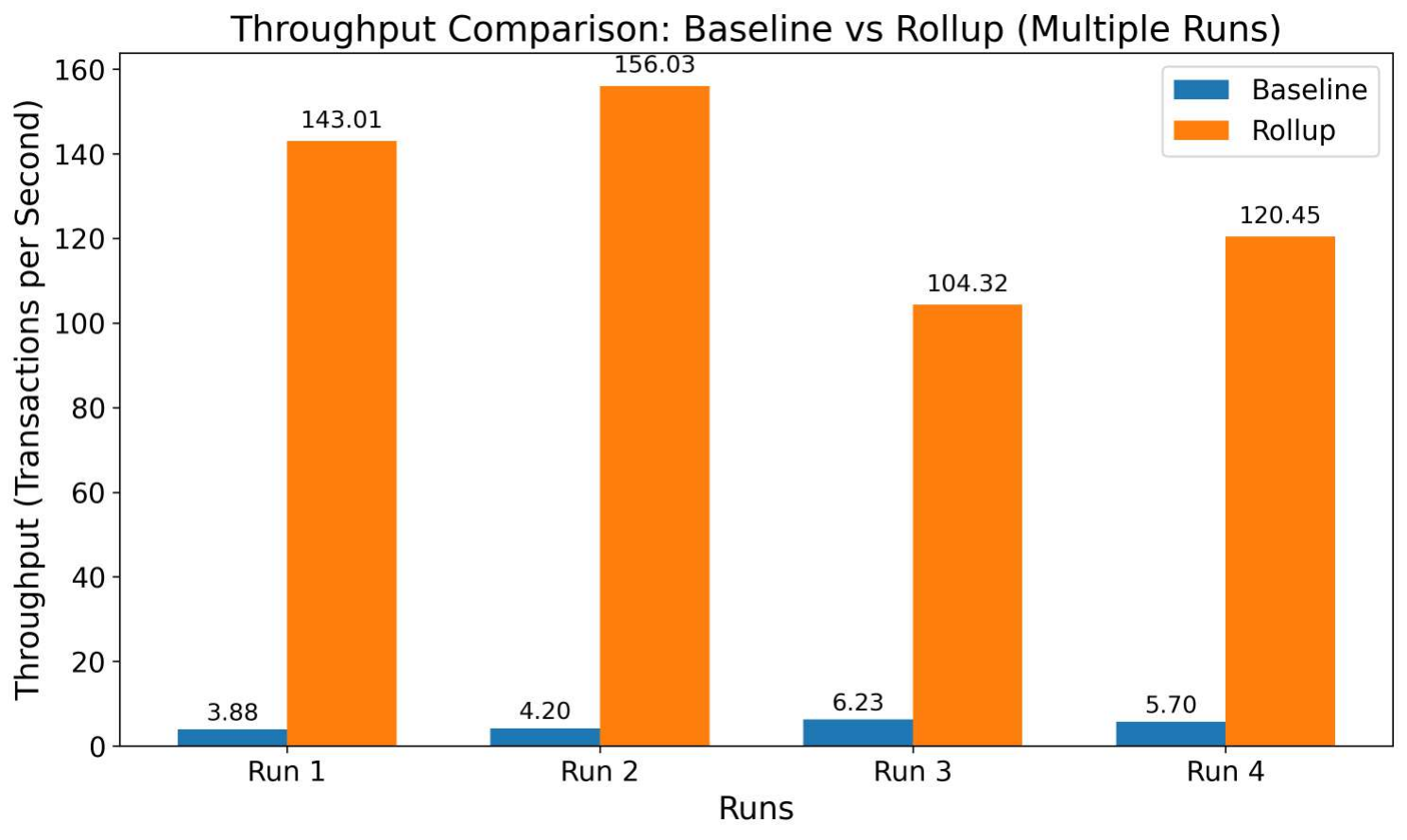}
    \caption{Ingestion throughput comparison across multiple iterations of the Baseline with the ZK-Rollup method}
    \label{fig:throughput}
\end{figure*}

\begin{figure*}[htbp]
    \centering
    \includegraphics[width=\textwidth]{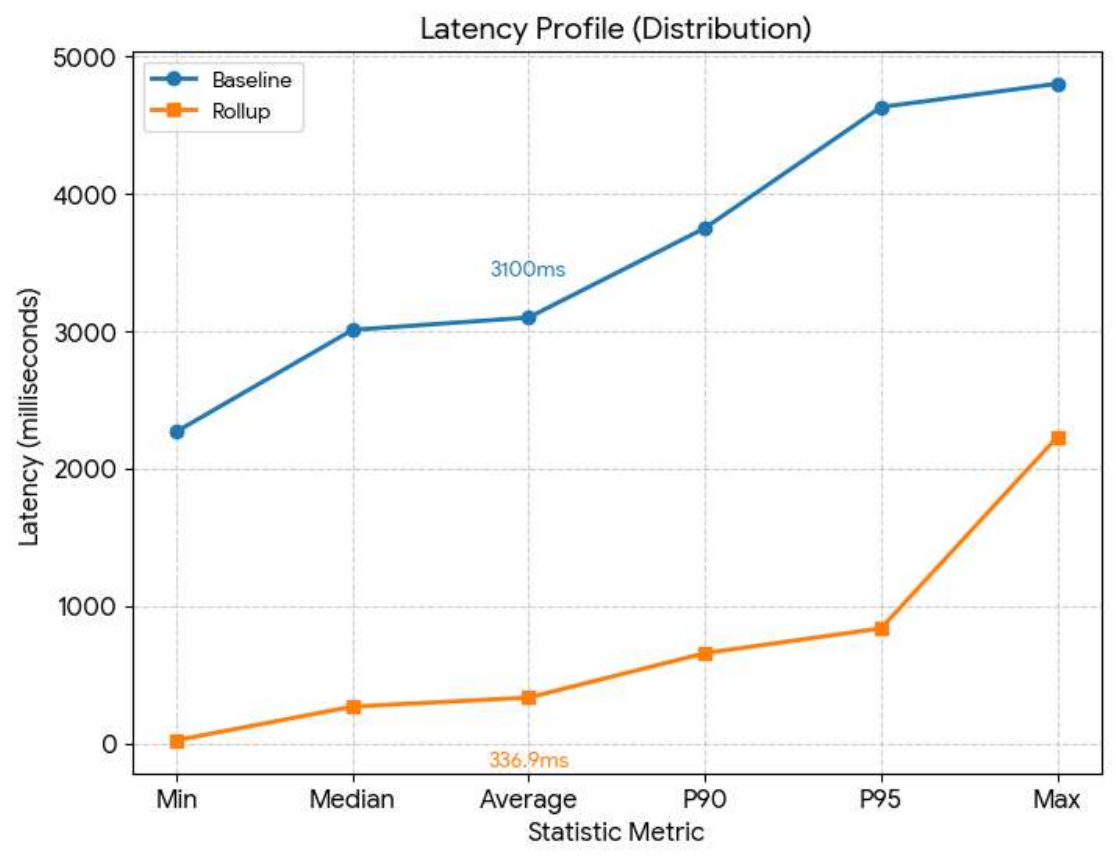}
    \caption{Latency distribution of various statistics of both Baseline model and the Rollup Model}
    \label{fig:latency}
\end{figure*}

Our ZK-Rollup method leads to a massive increase in throughput from 5-7 TPS from the baseline architecture to an ingestion rate of 100+ TPS (Figure~\ref{fig:throughput}). The baseline method is bailed out by blockchain consensus, making every client wait for a transaction to be finalized. The ZK-Rollup architecture decouples ingestion from settlement and instead handles around 100+TPS. Therefore, the architecture can absorb high demand spikes (for eg a Metaverse event) without failing. The client perceived latency also decreases massively due to the off-chain sequencer by nearly 10x, going from 2500-3500 ms to 300-400 ms (Figure~\ref{fig:latency}). This reduction in latency provides a responsive, real-time user experience.

\begin{figure*}[htbp]
    \centering
    \includegraphics[width=\textwidth]{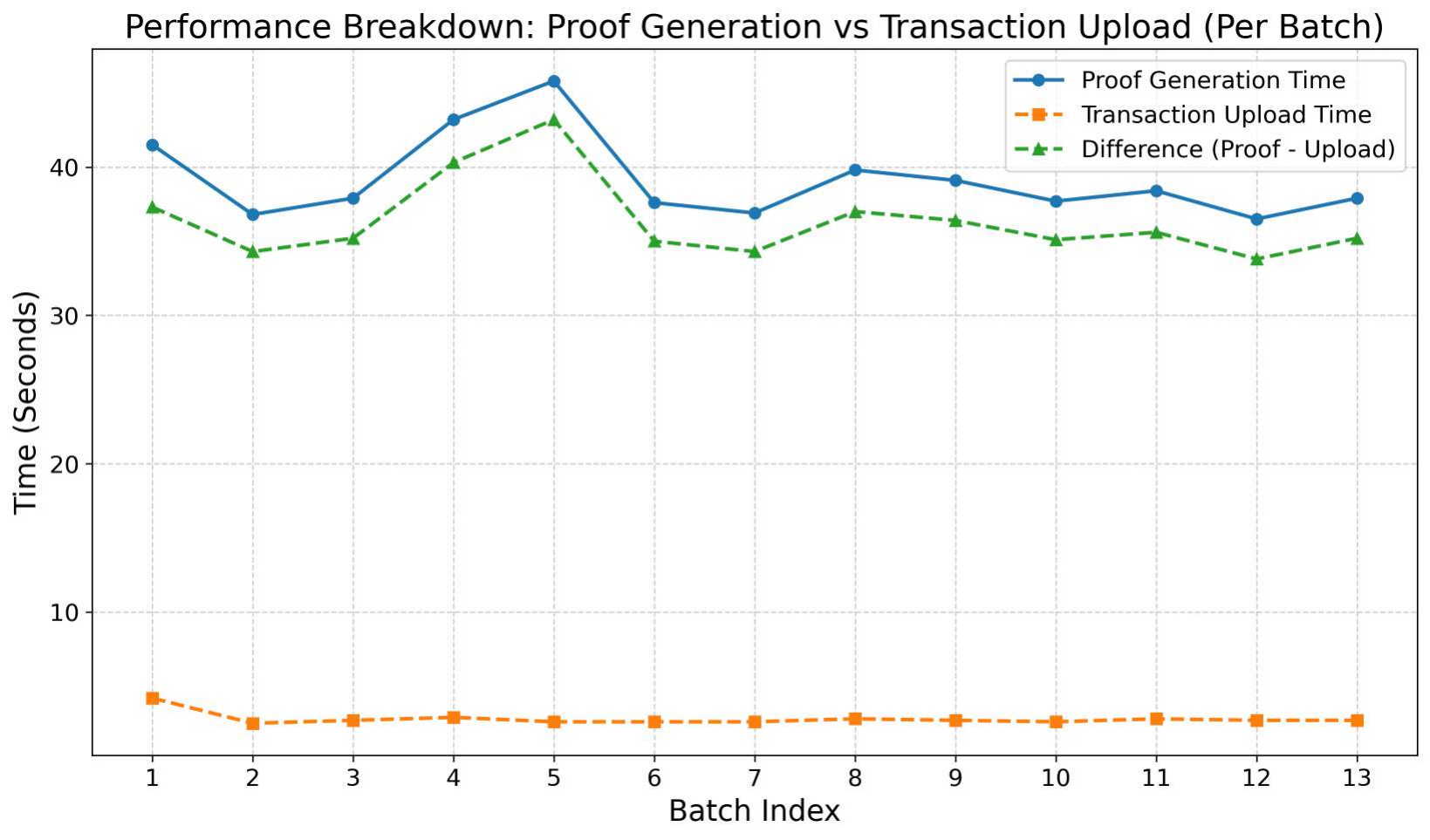}
    \caption{ZK-Proof generation time compared with transaction upload time for each batch}
    \label{fig:zk_time}
\end{figure*}

The ZK-Proof generation time takes nearly 35-45 seconds per batch due to the test hardware. On production hardware, this proof time would be negligible due to parallelization on high-performance servers (using GPUs or FPGA acceleration, generating proofs in <1 second), allowing the actual transaction completion throughput to also match the ingestion rate. The measured settlement throughput is strictly CPU-bound due to the proof generation time being a bottleneck. The transaction upload time is seen to be nearly negligible as well, with around 1-2 seconds taking to commit the batch to IPFS (Figure~\ref{fig:zk_time}). With production hardware, the settlement throughput would theoretically match the ingestion throughput of 100+ TPS, far exceeding the baseline.

\section{Related Work}

Table~\ref{tab:zkrollup_literature} presents a comparative summary of major works related to ZK-Rollups, and categorizes them based on whether they address the rollup mechanisms, performance evaluation, permissioned blockchain systems and address insights specific to Hyperledger Fabric architecture. As shown in the table, while a substantial amount of work exists on ZK-Rollups in public blockchains and their benchmarking, very few works address their relevance in permissioned blockchain systems. None of the works mentioned explore a Layer-2 ZK-Rollup architecture tailored specifically to Hyperledger Fabric, thus motivating the approach taken in the paper. 

\begin{table*}[!t]
\centering
\footnotesize
\caption{Comparison of Preexisting Work. $\checkmark$ indicates full support, $\triangle$ indicates partial support, and $\times$ indicates no support.}
\small
\makebox[\textwidth][c]{%
\begin{tabular}{|p{2.6cm}|p{2.2cm}|p{2.2cm}|p{2.2cm}|p{2.0cm}|p{4.0cm}|}
\hline
\textbf{Citation} &
\textbf{Full ZK-Rollup Architecture} &
\textbf{Performance Eval.} &
\textbf{Permissioned} &
\textbf{Fabric Insights} &
\textbf{Relevance} \\
\hline
Thibault et al. \cite{thibault2022blockchainscaling} & $\triangle$ & \texttimes & \texttimes & \texttimes & Foundational survey on rollup architectures \\
\hline
Katsika et al. \cite{katsika2025criticalrollups} & $\triangle$ & \texttimes & \texttimes & \texttimes & Conceptual comparison of ZK vs. Optimistic rollups \\
\hline
Chaliasos et al. \cite{chaliasos2024benchmarking} & \checkmark & \checkmark & \texttimes & \texttimes & System-level benchmarking of ZK-Rollups \\
\hline
Bruschi et al. \cite{bruschi2023achievements} & \texttimes & \texttimes & \texttimes & \texttimes & ZKP aggregation feasibility \\
\hline
Chen et al. \cite{chen2023zksync} & \texttimes & $\triangle$  & \texttimes & \texttimes & zkSync ecosystem analytics \\
\hline
Fekete \& Kiss \cite{fekete2023highereducation} & \texttimes & \texttimes & \texttimes & \texttimes & Smart-contract ZKP integration \\
\hline
Wang et al. \cite{wang2023supplychain} & $\triangle$ & \checkmark & \texttimes & \texttimes & Batched ZKP verification in supply chains \\
\hline
Pan et al. \cite{pan2024zkp} & \texttimes & \checkmark & \checkmark & \checkmark & Empirical ZKP performance analysis in Fabric \\
\hline
Yamamoto \& Yamashita \cite{yamamoto2023interoperability} & $\triangle$ & \checkmark & \texttimes & \texttimes & Highlights computational costs \\
\hline
Tortola et al. \cite{tortola2024libraries} & \texttimes & \checkmark & \texttimes & \texttimes & Comparative evaluation of zk-SNARK libraries \\
\hline
Ma \& Zhang \cite{ma2024healthcare} & $\triangle$ & \checkmark & \texttimes & \texttimes & Healthcare ZK-Rollup system with performance metrics \\
\hline
This work & \checkmark & \checkmark & \checkmark & \checkmark & ZK-Rollup framework for Hyperledger Fabric with empirical evaluation \\
\hline
\end{tabular}
}
\label{tab:zkrollup_literature}
\end{table*}

As summarized in Table~\ref{tab:zkrollup_literature}, existing work on ZK-Rollups and ZK-Proofs largely focuses on public blockchain systems and on cryptographic benchmarking, with limited attention to permissioned settings such as Hyperledger Fabric. A lot of early work under this topic has been found to focus on public blockchain systems where ZK-rollups have been found to drastically improve transaction throughput without compromising on security and reducing on-chain computation heavily, demonstrating that these can significantly scale decentralised applications~\cite{braun2018zkrollups}. Hyperledger Fabric has been found to impose strict limits on transaction throughputs due to consensus mechanisms like RAFT despite its permissioned nature offering modularity~\cite{vukolic2015quest}. The baseline performance observed in our experiments confirms these documented bottlenecks.

A significant contribution to the use of ZK-Proofs in Hyperledger Fabric has been the 2024 study by Pan et. al. ~\cite{pan2024zkp} which demonstrates that the implementation of ZK-Proofs in chain introduces a performance degradation of nearly 30-87.5\%. The challenge lies in aligning Fabric's endorsement policy with ZK-Proof verification mechanisms. Their results highlight how proof verification and generation (especially when on chain) create a bottleneck and significantly increase fabric’s processing time and latency.

The findings of Pan et al. are consistent with ZK-Rollup implementations in which the introduced computational overheads offset the expected gains from the introduced cryptographic loads\cite{yamamoto2023interoperability}. Efforts to reduce these costs are gaining momentum using more efficient ZK-SNARK libraries. Tortola et. al.\cite{tortola2024libraries} provide a comparative analysis of libraries used such as ZoKrates and Circom, helping in identifying tradeoffs in proof generation times and verification cost that become critical for practical deployment in permissioned settings. 

We see from the existing contributions to the field of ZK-Rollups that they offer a promising path forward for privacy preservation and scalability for blockchain networks despite their complexity. This is further corroborated by healthcare professionals that combine ZK-Rollups and decentralized storage systems like IPFS together with blockchain to protect patient data and also ensure system efficiency\cite{ma2024healthcare}. Existing studies focus on ZKP primitives or ZK-Rollups in public blockchains, whereas our work is the first to design and empirically evaluate a ZK-Rollup framework tailored for Hyperledger Fabric.”

\section{Conclusion}

This work proposed and implemented a ZK-Rollup architecture to address the scalability concerns in Hyperledger Fabric. By introducing an off-chain sequencer to help in decoupling transaction ingestion, we successfully address the inherent scalability limitations of standard Fabric architecture. Overall, this work demonstrates that ZK-Rollups are practical and highly beneficial in permissioned blockchain environments, paving the way for scalable enterprise deployments. Computational burden can be reduced on peer nodes and significantly boost transaction throughput while retaining the verifiability and confidentiality needed in enterprise and government-grade applications.

The primary limitation was found to be the proof generation time. On the test hardware (8 CPU cores), generating a ZK-SNARK proof took 35-45 seconds, becoming a big reason for the settlement throughput increasing and thus creating a bottleneck. As this settlement process is CPU bound, this architecture allows ingestion to proceed independently, ensuring the user latency is reduced.

Future work could include exploring the optimisation of this solution for enterprises, implementing this on high performance servers using GPUs or FPGAs to address the proof generation bottleneck. Future iterations of this work can also include dynamic batch sizing as the current work utilises fixed batches of 32 transactions to optimise the circuit, especially during periods of variable network load. Future works can also include testing different chaincode functions such as the transfer of assets between different organisations.

\section{Acknowledgement}

The authors would like to thank the Department of Computer Science, Birla Institute of Technology and Science, Pilani, for providing the computational resources and infrastructure required for this work. The authors would also like to acknowledge the use of ChatGPT in improving the presentation and grammar of the paper. The paper remains an accurate representation of the authors' underlying contributions.

\end{document}